\documentclass[12pt]{article}

\usepackage{times}
\usepackage{fixltx2e}
\usepackage{graphicx}
\usepackage[table]{xcolor}
\usepackage{ragged2e}
\usepackage{multirow}
\usepackage{helvet}
\usepackage{booktabs}
\usepackage{colortbl}
\usepackage{amsmath}

\usepackage{cmbright}

\usepackage{hyperref}
\hypersetup{
    colorlinks=true,
    linkcolor=sared,
    filecolor=magenta, 
    citecolor=sared,
    urlcolor=red,
}

\usepackage{blindtext}

\makeatletter
\newenvironment{figurehere}
{\def\@captype{figure}}
{}
\makeatletter

\usepackage{ccaption}
\usepackage{siunitx}
\usepackage{fixltx2e}
\newenvironment{sciabstract}{%
\begin{quote} \bf}
{\end{quote}}

\definecolor{sared}{rgb}{0.69, 0,0}

\newcounter{lastnote}

\title{\textbf{An In-Situ Spatial-Temporal Sequence Detector for Neuromorphic Vision Sensor
Empowered by High Density Vertical NAND Storage}}

\author
{Zijian Zhao$^{1\dag}$, Varun Darshana Parekh$^{2\dag}$, Po-Kai Hsu$^{3}$, Yixin Qin$^{1}$, \\
Yiming Song$^{1}$, A N M Nafiul Islam$^{2}$, Ningyuan Cao$^{1}$, Siddharth Joshi$^{1}$, \\ Thomas K{\"a}mpfe$^{4,5}$, Moonyoung Jung$^{6}$, Kwangyou Seo$^{6}$, Kwangsoo Kim$^{6}$, \\ Wanki Kim$^{6}$, Daewon Ha$^{6}$, Sourav Dutta$^{7}$, Abhronil Sengupta$^{2}$, Xiao Gong$^{8}$, \\ Shimeng Yu$^{3}$, Vijaykrishnan Narayanan$^{2}$, Kai Ni$^{1*}$
\\
\\
\normalsize{$^{1}$University of Notre Dame, Notre Dame, USA;}\\
\normalsize{$^{2}$Pennsylvania State University, State College, USA;}\\
\normalsize{$^{3}$Georgia Institute of Technology, Atlanta, USA;}\\
\normalsize{$^{4}$Fraunhofer IPMS, Dresden, Germany;}\\
\normalsize{$^{5}$TU Braunschweig, Braunschweig, Germany;}\\
\normalsize{$^{6}$Samsung Electronics Co., Ltd, South Korea;}\\
\normalsize{$^{7}$University of Texas, Dallas, USA;}\\
\normalsize{$^{8}$National University of Singapore, Singapore;}\\
\normalsize{$^\dag$Equal contribution;}\\
\normalsize{$^\ast$To whom correspondence should be addressed; E-mail:}\\ \normalsize{kni@nd.edu}
}

\date{}

\begin{document} 

\maketitle 
\begin{sciabstract}

Neuromorphic vision sensors require efficient real-time pattern recognition, yet conventional architectures struggle with energy and latency constraints. Here, we present a novel in-situ spatiotemporal sequence detector that leverages vertical NAND storage to achieve massively parallel pattern detection. By encoding each cell with two single-transistor-based multi-level cell (MLC) memory elements, such as ferroelectric field-effect transistors (FeFETs), and mapping a pixel’s temporal sequence onto consecutive word lines (WLs), we enable direct temporal pattern detection within NAND strings. Each NAND string serves as a dedicated reference for a single pixel, while different blocks store patterns for distinct pixels, allowing large-scale spatial-temporal pattern recognition via simple direct bit-line (BL) sensing, a well-established operation in vertical NAND storage. We experimentally validate our approach at both the cell and array levels, demonstrating that vertical NAND-based detector achieves more than six orders of magnitude improvement in energy efficiency and more than three orders of magnitude reduction in latency compared to conventional CPU-based methods. These findings establish vertical NAND storage as a scalable and energy-efficient solution for next-generation neuromorphic vision processing.

\end{sciabstract}

\section*{\textcolor{sared}{Introduction}}
\label{sec:introduction}
The unprecedented proliferation of neuromorphic sensing systems has exposed fundamental limitations in traditional von Neumann computing architectures, particularly in processing temporally-precise, event-driven data streams. This limitation is exemplified in neuromorphic vision sensors—event-based cameras, as shown in Fig.\ref{fig:implementation}(a), that generate sparse, asynchronous spike events in response to local pixel intensity changes, as illustrated in Fig.\ref{fig:implementation}(b). These sensors, characterized by sub-millisecond temporal resolution, 120 dB dynamic range, and power efficiency three orders of magnitude better than conventional cameras, are transforming capabilities in robotics and computer vision \cite{Gallego_2022}. The precise temporal data streams from these sensors enable advances in brain-computer interfaces, cognitive robotics \cite{Kasabov2012}, and autonomous systems \cite{Devulapally_2024_CVPR}; however, their full potential remains constrained by the current computing architectures that separate sensing, processing, and storage components.
Extracting meaningful information from these temporally-precise data streams requires sophisticated spatiotemporal pattern matching (STPM) techniques, which have emerged as essential tools for analyzing complex temporal phenomena. These techniques have enabled fundamental discoveries across multiple scientific domains: in neuroscience, revealing previously unobservable dynamics in cortical circuits and epileptic activity \cite{KOHLING200217}; in healthcare, enabling precise cardiac diagnostics \cite{Syeda2007Characterizing} and advancing our understanding of language processing in aging and aphasia \cite{Wang2023, Kries2024.10.21.619562}; and in environmental science, detecting ocean eddy formations and quantifying climate change impacts \cite{BERMINGHAM2019334, 10.1145/191246.191296}. The application of STPM to video analysis has advanced sequence alignment techniques \cite{Caspi2002}, while in surveillance systems, it has enabled real-time pattern recognition \cite{Zhang2013Survey}.


The conventional Von Neumann hardware implementation for processing sensor data streams suffer from inherent bottlenecks due to frequent data movement among sensor, memory, and processing units, thus imposing fundamental limits on latency and energy efficiency \cite{Valencia19}. Moreover, current hardware implementations also require complex analog-to-digital conversion chains and face intrinsic limitations in real-time processing capability. Optical computing methods that utilize geometric shape matching algorithms achieve high accuracy \cite{Gook18}, but their reliance on centralized processing makes them unsuitable for edge applications. Field-programmable gate array (FPGA) implementations of spiking neural networks \cite{Caron11}, despite their inherent parallelism, encounter scalability constraints with high-dimensional temporal data \cite{Schaik24}. Photonic neural networks show promise in ultra-fast pattern recognition \cite{Han23} but face significant challenges in weight programming and temporal processing. Probabilistic approaches using Hidden Markov Models effectively handle spatial uncertainty \cite{BERMINGHAM2019334} but require substantial computational resources that preclude real-time processing of large-scale temporal datasets. Recent advances in compute-in-memory (CiM) architectures have demonstrated promising approaches to minimize data movement through in-memory computation. 
Several CiM hardware solutions, such as content-addressable memory (CAM), have been proposed for pattern matching, where the search pattern is compared with stored patterns in a massively parallel fashion. Leveraging this capability, various machine learning applications, including augmented associative memory, have been explored \cite{ni2019ferroelectric, hu2021memory}. Additionally, emerging memory technologies have been introduced to enhance CAM density, even with single-memory-based CAM cell designs \cite{Wang24SingleFefet, luo2022novel, zhang2023ultra, yin2023ultracompact}. However, these CiM solutions continue to face significant challenges in processing temporal data streams, as they require complex peripheral circuitry for temporal correlation and lack direct sensor interfaces.

Here, we introduce an in-situ spatiotemporal sequence detector utilizing the mature vertical NAND storage technology, as illustrated in Fig. \ref{fig:implementation}(c). The vertical NAND structure, consisting of single-transistor memory cells connected in series, is ideally suited for sequence detection. When appropriately mapped, it naturally performs sequence detection and achieves massive parallelism due to its ultrahigh memory density. In this array, \textit{N} word lines (WLs) are connected in series, with a transistor memory cell located at each intersection between a WL and the memory string. The bit line (BL) runs atop each string. A key advantage of vertical NAND is its ability to support a large number of BLs (e.g., over 10,000), enabling the exploitation of massive parallelism for sequence detection. Fig. \ref{fig:implementation}(d) depicts the mapping of reference sequence patterns and incoming pulse sequences to the vertical NAND array. In this configuration, one block stores the reference patterns for each pixel, while each string holds reference pulse patterns across all time steps. Incoming pulse patterns are mapped to WL pulses, with the corresponding WL determined by the pulse sequence. For instance, as shown in Fig. \ref{fig:implementation}(e), the reference pattern is stored by programming two consecutive devices to encode one time step of the reference pattern. This allows \textit{N} WLs to store \textit{N}/2 time steps. The incoming pulse patterns are then mapped to the WL reading voltages of the cells. The pulse at time step 1 corresponds to the WL voltage of the bottom cell and time step 2 maps to the WL voltage of the cell on top of the bottom cell. With this mapping, when the incoming pulses for each pixel matches the stored temporal pattern, all the transistors in the string will be turned ON and the corresponding BL current will be high, as demonstrated in Fig. \ref{fig:implementation}(f). This shows that the vertical NAND array, without any structure modification, naturally implements a spatial-temporal sequence detector.   


\begin{figurehere}
\centering
\includegraphics[width=0.9\linewidth]{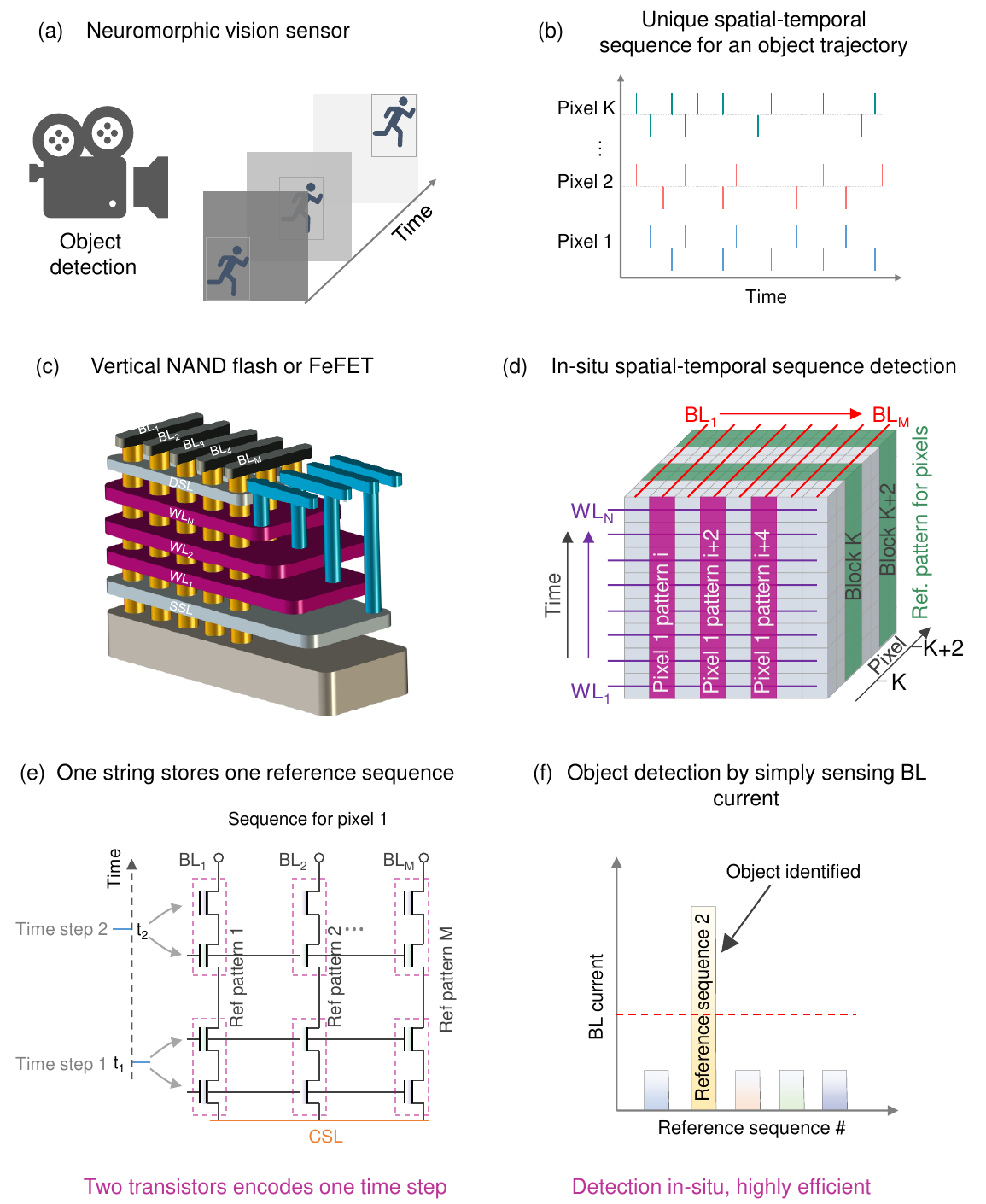}
\caption{\textit{\textbf{Implementation architecture of near-sensor processing.} (a) Neuromorphic vision sensor generates (b) precise spatial-temporal sequences processed through (c) 3D NAND array utilizing (d) novel mapping schemes and (e) temporal encoding. (f) Pattern detection is achieved through bit-line current sensing.}}
\label{fig:implementation}
\end{figurehere}

This work introduces a processing-near-sensor-and-storage architecture that fundamentally redefines the integration of sensing, computation, and memory storage. By eliminating the von Neumann bottleneck and enabling massively parallel temporal processing, our system achieves a three-order-of-magnitude improvement in energy efficiency and processing speed compared to conventional architectures. Unlike previous NAND-based sequence filters \cite{chen2023multi}—which repeatedly programs memory for each incoming pulse, leading to severe endurance limitations and the inability to store or directly compare patterns—our proposed design directly stores complete reference patterns. This approach allows robust and repeated spatial-temporal comparisons without endurance degradation, fundamentally overcoming previous challenges.

Importantly, our vertical NAND-based architecture extends significantly beyond sensor-centric workloads, providing transformative advantages for spatio-temporal pattern matching in large-scale databases and structured storage systems. Applications such as financial market analysis, genomic data queries, extensive text searches, anomaly detection in large-scale IoT networks, and real-time analysis of social media streams underline the growing need for scalable and low-latency database-centric pattern matching solutions~\cite{Agrawal2008SIGMOD,Zhu2023PVLDB}. Recent surveys have highlighted that traditional database methods struggle with the complexity and scalability of spatial and temporal queries, especially in the SQL ecosystem and deep learning-based spatio-temporal applications~\cite{Jitkajornwanich2020,Giatrakos2020VLDBJ}. Moreover, exact pattern matching approaches used in text and database searches often suffer from inefficiencies in memory consumption and query latency when scaled~\cite{Hakak2018}. By integrating computation directly within 3D-stacked NAND storage, our approach dramatically reduces data movement, leverages intrinsic parallelism, and thus resolves key performance bottlenecks inherent in conventional systems. Ultimately, our architecture represents a significant advancement towards real-time spatio-temporal analytics at unprecedented scales, effectively bridging sensing, storage, and database processing.


\section*{\textcolor{sared}{Device Design and Working Principles}}
\label{sec:principle}

The proposed solution is generally applicable to any transistor memory that can be integrated in the vertical NAND architecture, such as charge trapping based flash or ferroelectric field effect transistor (FeFET). 
Without loss of generality, this work adopts FeFET NAND array for demonstration. HfO\textsubscript{2} based FeFET has recently seen great interests in its application to enable the next generation of vertical NAND storage, due to its lower write voltage and fast programming speed \cite{yoon2023qlc, lim2023comprehensive, myeong2024strategies, kim2024depth, zhao2024large, qin2024}. A major roadblock to its application in vertical NAND storage is that its memory window is linearly proportional to the ferroelectric thickness, thus limiting the string density if a memory window comparable to flash is needed for multi bit storage \cite{zhao2024large, qin2024}. Recent breakthroughs that lift such a bottleneck have been reported by exploiting gate-side charge injection that can boost the memory window alongside the polarization defined window \cite{yoon2023qlc, lim2023comprehensive, myeong2024strategies, kim2024depth, zhao2024large, qin2024}. Active research are ongoing to address various process integration issues, reliability optimization, and performance improvement. Orthogonal to those exciting developments, this work uses a baseline FeFET NAND array for the demonstration. 

\begin{figurehere}
	\centering
	\includegraphics[width=0.95\linewidth]{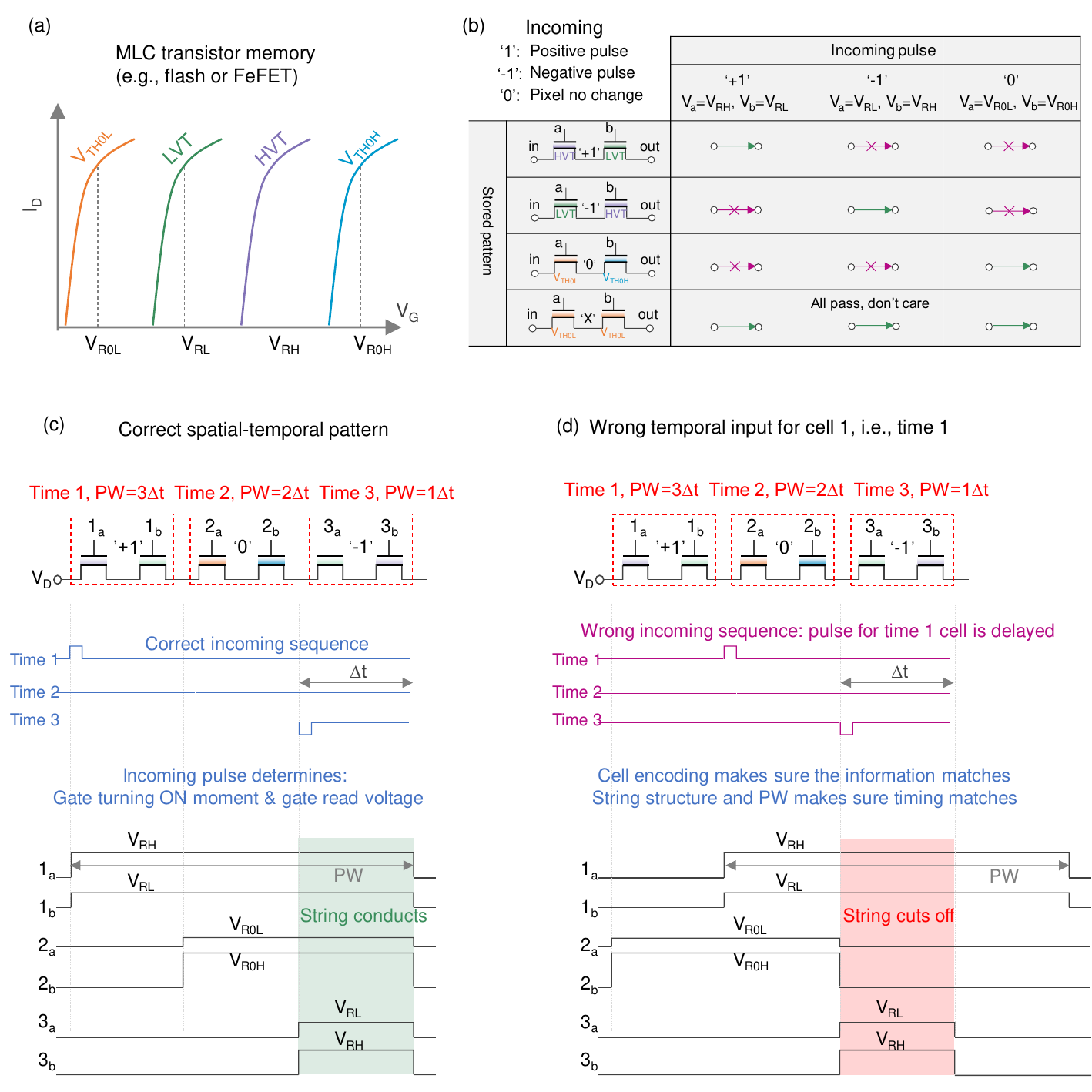}
	\caption{\textit{\textbf{Working principles of the spatial-temporal sequence detection.} (a) 4-level MLC is required to encode all the four states. (b) Corresponding encoding for the stored patterns and incoming pulses such that correct functionality is achieved. (c) With matching spatial patterns and correct incoming temporal sequences, the string is turned ON and produces a high BL current. (d) With wrong incoming temporal sequences, the string remains OFF even matching spatial patterns are applied.}}
	\label{fig:encoding}
\end{figurehere}

To encode the different states of the pixels, i.e., positive (‘+1’), negative (‘-1’), no change (‘0’), or don’t care (‘X’) for noisy pixel masking, a cell composed of two MLC FeFETs are adopted. The first 3 states correspond to the typical outputs of a pixel. The don't care state adds flexibility in focusing on the regions of interest and blocks the other areas. Fig.\ref{fig:encoding}(a) shows a typical \textit{I}\textsubscript{D}-\textit{V}\textsubscript{G} curve of a non-volatile MLC transistor memory. Four states of the MLC are noted as \textit{V}\textsubscript{TH0L}, low-\textit{V}\textsubscript{TH} (LVT), high-\textit{V}\textsubscript{TH} (HVT), \textit{V}\textsubscript{TH0H}, with corresponding read voltages, \textit{V}\textsubscript{R0L}, \textit{V}\textsubscript{RL}, \textit{V}\textsubscript{RH}, and \textit{V}\textsubscript{R0H}. The stored pattern is encoded as \textit{V}\textsubscript{TH} states of the two FeFETs. For example, as shown in Fig.\ref{fig:encoding}(b), stored '+1' state is encoded as HVT/LVT for cell a/b, respectively and stored '-1' state is represented as LVT/HVT for the cell a/b, respectively. In order to store the '0' state, an additional level is needed. In this work, we adopt the multi-level cell so that the additional \textit{V}\textsubscript{TH0L} and \textit{V}\textsubscript{TH0H} states are used to map the stored '0' state. Since don't care state need to pass signal irrespective of the incoming pulse, it can be realized by programming the two FeFETs to the lowest \textit{V}\textsubscript{TH} state, i.e., \textit{V}\textsubscript{TH0L}.  After the patterns are stored, the incoming pulses are encoded as the read voltages for the two FeFETs, as seen in Fig.\ref{fig:encoding}(b), such that the input can pass to the output only when incoming pulse matches stored ‘+1’, ‘-1’, or ‘0’ or simply pass when the ‘X’ is stored, as explained in Fig.\ref{fig:encoding}(b). For example, when the incoming pulse is a positive pulse ('+1'), the gate voltages applied to the two FeFETs are \textit{V}\textsubscript{RH} and \textit{V}\textsubscript{RL}, respectively. If the stored pattern is '+1', which means that the two FeFETs are written to the HVT state and LVT state, both FeFETs are turned on, and hence the input can pass through to the output. 
In the case the stored pattern is ‘-1’ or '0', i.e., mismatch, the FeFET in HVT state will block the signal transmission, thus realizing the desired functionality. This mapping outperforms the previously proposed design \cite{chen2023multi}, where each incoming pulse requires programming of the memory and thus will suffer from the limited endurance. In contrast, in the proposed approach, memory states are fixed and simple read operations are performed, avoiding the reliability issue.

Built upon the two transistor cell that can perform pattern matching, the temporal sequence detection can be realized in the string with serially connected cells. 
The string naturally enforces temporal order, as it can only conduct from one end to the other. As discussed in Fig. \ref{fig:implementation}(e), if the bottom cell encodes time step \textit{n} and the top cell encodes time step \textit{n}+1, then pattern matching at step \textit{n}+1 can only occur when the time step n matches. Consequently, a string can fully pass the input signal only if all time steps match sequentially. A key aspect of implementation is the timing diagram of the WL voltages. Two mappings can be adopted. In one approach, each WL receives a voltage pulse with the same pulse width. Each incoming pulse either turns the cell ON or OFF for a unit pulse width. If a pattern match occurs, the bottom cell’s charge is transferred upward. This design is analogous to a charge-coupled device (CCD), where charge is sequentially transferred from one stage to another. If, at the end, the charge packet is successfully transferred to the BL, the sequence is fully matched. This design is intuitive and attractive, but faces several challenges. In vertical NAND storage, each transistor has a small gate length and the charge packet is small. The charge loss during the charge transfer could significantly degrade the sense margin. This issue will get worse with further vertical scaling of vertical NAND. Alternatively, this work adopts the second approach, which utilizes the whole string conduction for detection, which offers high sense margin and scalability. In this design, varying pulse widths are assigned to each cell and that the bottom cells receive larger pulse widths than the top cells such that when sequence matching happens, there will be a time duration that all the cells in the string are conducting simultaneously that current can flow from bottom to up. Specifically, to ensure detection of the correct temporal order in a string with \textit{N} pixel cells (i.e., 2\textit{N} FeFETs), each cell (with index \textit{i} counted from the input side) is applied with gate pulses with a width of $(N+1-i)\Delta t$, where $\Delta t$ is the unit pulse width. The application of the gate pulses is triggered by the spike and the pulse amplitudes are determined by the spike value, as explained in Fig.\ref{fig:encoding}(b). Fig.\ref{fig:encoding}(c) and (d) illustrate two in-situ spatio-temporal pattern detections implemented by NAND string structures. Fig.\ref{fig:encoding}(c) illustrates a scenario where both spatial pattern and temporal sequence match. During time step 3, all FeFETs are turned on and thus the NAND string can pass the input to the output. In the other case where the temporal sequence does not match, even if the incoming pulses are the same, the string is not conductive with wrong temporal sequence, as illustrated in Fig.\ref{fig:encoding}(d).

\section*{\textcolor{sared}{Experimental Verification}}

For proof-of-concept, experimental verification is implemented using Hf\textsubscript{0.5}Zr\textsubscript{0.5}O\textsubscript{2} based FeFET technology, which is compatible with vertical NAND architecture. 
In this work, basic FeFET without introducing gate side injection is adopted. To have a enough memory window to realize the required MLC, a 20nm thick ferroelectric layer is used. But to reduce the formation of the monoclinic dielectric phase in the thick Hf\textsubscript{0.5}Zr\textsubscript{0.5}O\textsubscript{2} film, a laminated thin film is used where a thin Al\textsubscript{2}O\textsubscript{3} layer (1 nm) is inserted after 10-nm Hf\textsubscript{0.5}Zr\textsubscript{0.5}O\textsubscript{2} growth, followed by another 10-nm Hf\textsubscript{0.5}Zr\textsubscript{0.5}O\textsubscript{2} growth \cite{kim2014grain, qin2024understanding}. The SEM image of the NAND string is shown in Fig.\ref{fig:experiment}(a), which is composed of 5 FeFETs connected in series. The schematic cross section of the gate stack and the transmission electron microscopy (TEM) cross section are shown in  Fig.\ref{fig:experiment}(b) and Fig.\ref{fig:experiment}(c), respectively. It clearly shows that the 1nm Al\textsubscript{2}O\textsubscript{3} separates the two Hf\textsubscript{0.5}Zr\textsubscript{0.5}O\textsubscript{2} layer and the grain growth in Hf\textsubscript{0.5}Zr\textsubscript{0.5}O\textsubscript{2} layers is interrupted by the Al\textsubscript{2}O\textsubscript{3}, thus suppressing the monoclinic phase stabilization and favoring the ferroelectric phase. The elemental map shown in Supplementary Fig.\ref{fig_s1} confirms the elements in the gate stack. 

Fig.\ref{fig:experiment}(d) shows that with a 20nm thick Hf\textsubscript{0.5}Zr\textsubscript{0.5}O\textsubscript{2} film, a memory window (MW) larger than 4 \textit{V} is realized. As a result, 4 evenly distributed threshold voltage (\textit{V}\textsubscript{TH}) states can be accommodated. realizing a multi-level cell (MLC). Eight devices are presented, showing reasonable device variability. The FeFET MLC \textit{V}\textsubscript{TH} histogram and the cumulative probability distribution are shown in Supplementary Fig.\ref{fig_s2}. 
With the test structure, first the single cell operation is verified. Fig.\ref{fig:experiment}(e) shows the measured cell current for all the 12 cases as explained in the logic truth table shown in Fig.\ref{fig:encoding}(b). Four different patterns, i.e., '+1', '-1', '-0', and 'X' are stored in the cell by programming the two FeFETs into corresponding \textit{V}\textsubscript{TH} states. Then gate pulses corresponding to 3 different incoming pulses, i.e., '+1', '-1', and '0' are then applied to the cell and the output bit line current is measured. These results clearly validate the truth table and show that only when the incoming pulse matches the stored pattern, the output current is high as both FeFETs are turned ON. Any other scenario will result in a low current at the noise floor as at least one FeFET in the cell is cut off. As a result, there is a large margin between the match and the mismatch scenarioes, showing the effectiveness of the proposed design.


\begin{figurehere}
	\centering
	\includegraphics[width=0.95\linewidth]{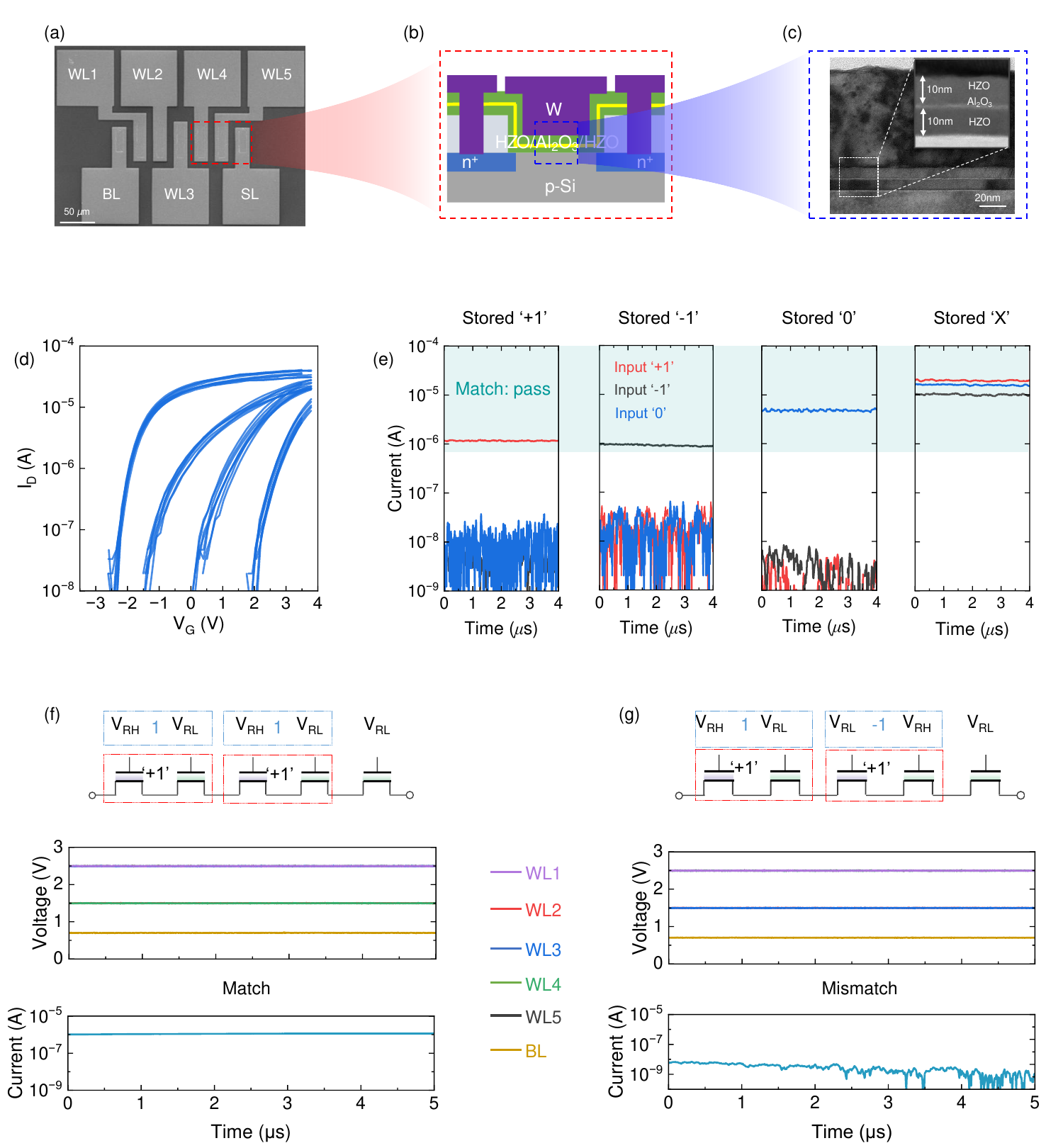}
	\caption{\textit{\textbf{Experimental verification with 5-FeFET NAND string.} (a) SEM of the 5-FeFET NAND string. (b) Cross-section of the developed FeFET. (c) TEM of the gate stack. 20 nm HZO is interrupted with 1 nm Al\textsubscript{2}O\textsubscript{3}. (d) \textit{I}\textsubscript{D}-\textit{V}\textsubscript{G} curves showing MLC capability. (e) Readout current from a single FeFET for the four stored states with four inputs applied to each state. Only matching patterns produce a high BL current. State 'X' exhibits high BL current no matter what input is applied. (f) Readout current from the 5-FeFET NAND string for the case of match condition. A high BL current is observed. (g) In the case of mismatch condition, a low BL current is obtained.}}
	\label{fig:experiment}
\end{figurehere}

Using a 5-FeFET NAND string, the functionalities of the proposed spatio-temporal sequence detection are experimentally validated. Two examples of the current response in the final read cycle of the spatio-temporal sequence detection are shown in Fig.\ref{fig:experiment}(f) and (g). The stored patterns for all cells in both examples are '+1'. The fifth FeFET is used as a pass transistor and is always written to low-\textit{V}\textsubscript{TH} state. In Fig.\ref{fig:experiment}(f), the incoming pulses for the two cells are '+1' and '+1', respectively. Since the incoming pulses match the stored pattern, a high current is shown through the NAND string. However, if the incoming pulses ('+1' and '-1') do not match the stored patterns ('+1' and '+1'), the string exhibits a low current, as shown in Fig.\ref{fig:experiment}(g). Other scenarios are shown in Supplementary Fig.\ref{fig_s3}. All combinations of the stored spatial patterns and input temporal sequence are measured, demonstrating successful spatio-temporal sequence detection operations. 
\section*{\textcolor{sared}{Benchmarking}}




To evaluate the performance of the proposed design, we generated a dense spatiotemporal dataset consisting of cross ('×') and plus ('+') patterns across 10 time steps on an 8$\times$8 pixel grid using the Brian2 simulator \cite{Stimberg2019}. The dataset was synthesized with a leaky integrate-and-fire (LIF) neuron model \cite{neuronal14}, configured with a membrane time constant of 5ms and a threshold of 0.85V. The '×' pattern activates diagonal pixels, whereas the '+' pattern triggers events along the vertical and horizontal axes, with all other pixels masked to maintain shape fidelity. A total of 500 reference patterns were stored for detection tasks.


Benchmarking was conducted using two distinct algorithms: a brute-force sequential search and a Jaccard-based locality-sensitive hashing (LSH) method. In the brute-force approach, each query is compared exhaustively against all stored patterns. In contrast, the LSH method computes and compares compact hash representations to pre-filter candidate matches before exact evaluation. Both algorithms were implemented and profiled on an Intel 4-core CPU with 16GB DRAM and a 12MB cache. System-level performance—including execution latency, memory usage, and energy consumption—was measured using Intel VTune, the Python line profiler \cite{line_profiler}, and PyRAPL \cite{pyRAPL}. A single query refers to the evaluation of one full 10-timestep input sequence against all stored patterns. The 20-query scenario involves processing 20 such input sequences sequentially to assess scalability under heavier workloads.

\begin{figurehere}
	\centering
	\includegraphics[width=\linewidth]{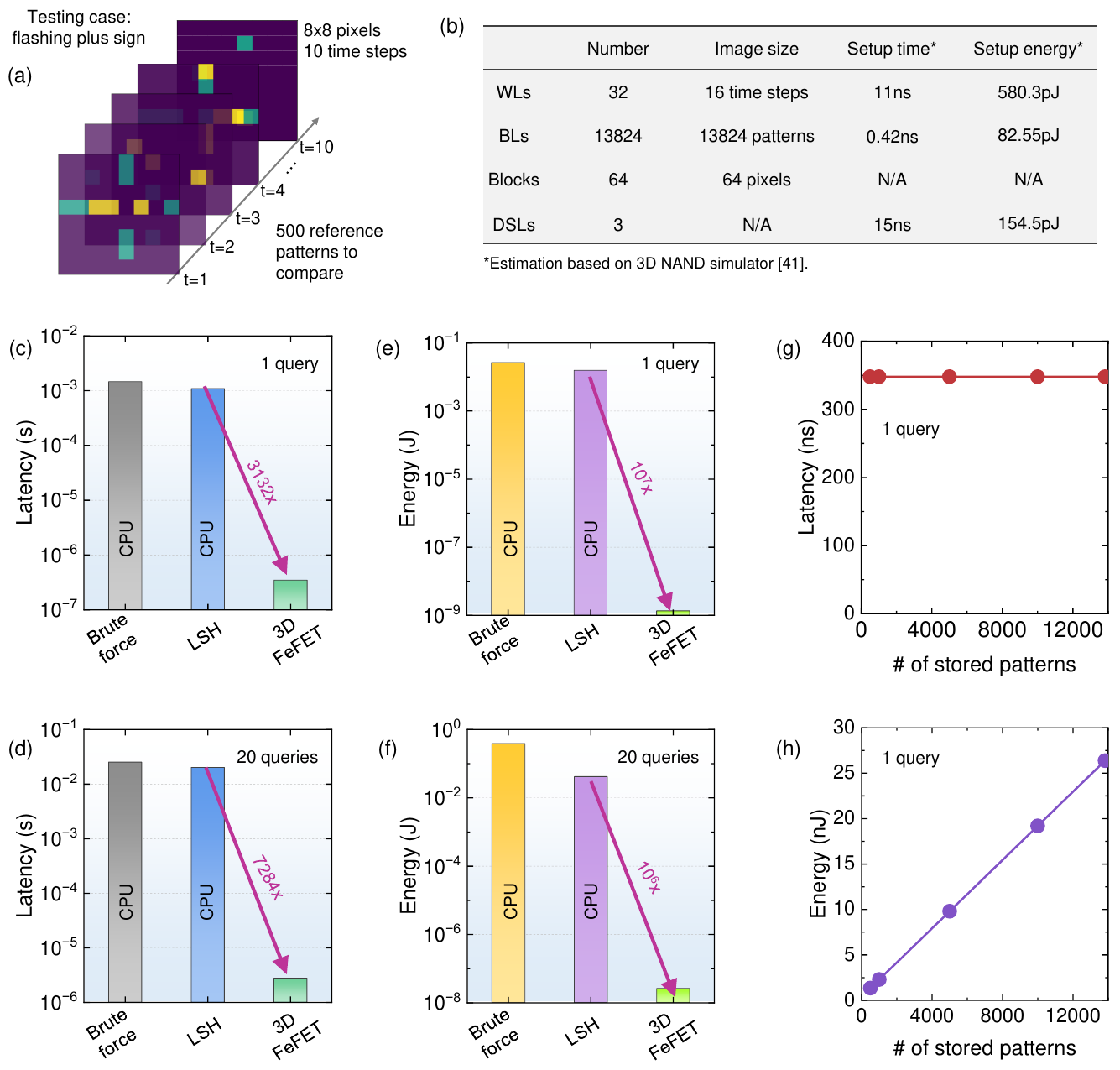}
	\caption{\textit{(a) A case of flashing plus/cross sign across 10 steps is tested. (b) 3D NAND simulation design \cite{Shim21}. (c)/(d) and (e)/(f) latency and energy to detect 1/20 incoming sequences, respectively. (g) Due to massive parallelism, query latency is constant. This is because all BLs are activated simultaneously, causing the same delay. All patterns fit into blocks. (h) query energy consumption grows linearly with the number of bit lines and hence the number of stored patterns
}}
	\label{fig: result}
\end{figurehere}

For the evaluation of 3D vertical NAND array in performing spatial-temporal sequence detection, a dedicated simulator is adopted \cite{Shim21}. Fig. \ref{fig: result}(b) shows the adopted design of the 3D NAND subarray for FeFET in the size of 64 blocks $\times$ 3 DSL $\times$ 32 WL $\times$ 13824 BL. Here the DSL is the drain select line in the vertical NAND to select the string. Unlike the previous work \cite{Lue19} of 16 DSL, the scaled subarray has a lighter RC latency for BL and WL due to smaller parasitics. Furthermore, the 3D FeFET has lower operation voltages compared to conventional charge-trapping based flash devices, which reduces the energy consumption. The calibrated latency and energy consumption are summarized in Fig.\ref{fig: result}(b). Based on the calibrated performance for 3D FeFET NAND subarray, the resulting latency/energy to execute the task are also projected for 1 query, shown in Fig.\ref{fig: result}(c)/(e), and 20 queries, shown in Fig.\ref{fig: result}(d)/(f), respectively.

Due to its in-memory computing nature and massive parallelism, the vertical NAND FeFET array shows more than 10\textsuperscript{3}$\times$/10\textsuperscript{6}$\times$ improvement over CPU baseline in latency/energy, respectively. In addition, its huge memory capacity and a large amount of BLs lead to an almost constant latency with the number of stored patterns, as shown in Fig.\ref{fig: result}(g), as long as those patterns can be stored in memory. In addition, the energy increases linearly with the number of stored patterns, as shown in Fig.\ref{fig: result}(h). These results highlight the efficacy of vertical NAND FeFET arrays for high-throughput, low-energy spatiotemporal sequence detection.
\newcommand{\mycomment}[1]{}

\section*{\textcolor{sared}{Conclusion}}
\label{sec:conclusion}

In summary, We have proposed an in-situ spatial-temporal sequence detector for neuromorphic vision sensor. Such a functionality is realized by exploiting the two MLC FeFETs cell and the intrinsic NAND string structure. With its extreme high memory density, massive parallelism, and superior energy efficiency, vertical NAND array is a highly promising platform for processing spatial-temporal signals.  

\mycomment{

\section*{\textcolor{sared}{Materials and Methods}}
\label{sec:materials}

\subsection*{Material characterization}
The microstructure of the ferroelectric layer is analyzed using HAADF 4D-STEM electron diffraction as well as transmission Kikuchi diffraction (TKD). STEM is carried out with a dedicated detector with high dynamic range at each pixel for automated crystal orientation mapping (ACOM). ACOM indexation has been achieved using the ASTAR software package. The TKD measurements are carried out using a Bruker Optimus detector head mounted into a scanning electron microscope. The acceleration voltage is 30 kV and a current of 3.2 nA is used. Similar to TKD, which has been utilized to investigate the granular structure of polycrystalline HfO\textsubscript{2} films, ACOM is carried out in transmission, with the detector located inside the beam. The main differences here are the accelerator voltage and the adjustable convergence angle, which allows for tuning the measured signal between classical electron diffraction and Kikuchi diffraction patterns.

\subsection*{Device fabrication}
FeFETs tested in this work are integrated on the 28 nm industrial high-$\kappa$ metal gate (HKMG) platform. The device has a gate stack composed of a poly-crystalline Si/TiN (2 nm)/doped HfO\textsubscript{2} (8 nm)/SiO\textsubscript{2} (1 nm)/p-Si. The ferroelectric gate stack process starts with an 8 nm thick doped HfO\textsubscript{2} deposition through atomic layer deposition process on the 300 mm silicon wafer, which is covered with a thin SiO\textsubscript{2} interfacial layer. Then a TiN metal gate electrode was deposited using physical vapor deposition (PVD), followed by the poly-Si gate electrode deposition. The source and drain n+ regions were obtained by phosphorous ion implantation, which were then activated by a rapid thermal annealing (RTA) at approximately 1000 $^\circ$C. In addition to the dopant activation, the RTA process will also stabilize the ferroelectric orthorhombic phase within the doped HfO\textsubscript{2}.

\subsection*{Electrical characterization}
The FeFET device characterization was performed with a PXI-Express system from National Instruments, using a PXIe-1095 cassis, NI PXIe-8880 controller, NI PXIe-6570 pin parametric measurement unit (PPMU) and NI PXIe-4143 source measure unit (SMU).
Prior to characterization all FeFETs are preconditioned using the SMUs by cycling them 100 times with the pulses of +4.5 V, -5 V with a pulse length of 500 ns each. Read out of the memory state is done by a step wise increase of the gate voltage in 0.1 V increments while applying 0.1 V to the drain terminal and measuring the current using the PPMU. Bulk and source terminals are tied to ground at all times. The read operation takes approximately 7 ms. The multi-level characterization of individual FeFETs is performed by putting them in a reference state with a gate voltage of -5 V or +4.5 V for 500 ns for erase or program, respectively. After that a single pulse of increasing amplitude is applied for 200 ns. The gate voltage amplitude stepping is set to 100 mV. After each pulse a delay of 2 s is added to ensure sufficient time for charge detrapping after which a readout is performed. For the characterization of a CAM word with a word length of 8, all the integrated FeFETs are wired together and then the total drain current is then measured to verify the array operation.
}

\section*{\large Data availability}
All data that support the findings of this study are included in the article and the Supplementary Information file. These data are available from the corresponding author upon request.

\section*{\large Code availability}
All the codes that support the findings of this study are available from the corresponding author upon request.

\bibliography{ref.bib}

\bibliographystyle{Nature}

\section*{\large Acknowledgments}

This work is supported by SUPREME and PRISM centers, two of the SRC/DARPA JUMP 2.0 centers and NSF 2344819 and 2346953.

\section*{\large Author contributions}

K.N. and V.N. proposed the project. Z.Z., Y.Q., and Y.S. conducted device integration and experimental verification. V.D.P., A.N.M.N.I., S.D., and A.S. performed application-level analysis. P.K.H. and S.Y. conducted array level benchmarking. N.C., S.J., and T.K. helped with mapping the function to the NAND array. M.J., K.S., K.K., W.K., D.H., and X.G. helped with the discussion of vertical NAND array. All authors contributed to write up of the manuscript.

\section*{\large Competing interests}
The authors declare that they have no competing interests.

\newpage
\renewcommand{\thefigure}{S\arabic{figure}}
\renewcommand{\thetable}{S\arabic{table}}

\onecolumn
\centering
\textbf{\Large Supplementary Information}
\setcounter{figure}{0}
\setcounter{table}{0}
\setcounter{page}{1}

\begin{flushleft} 
\textbf{\large Process Integration of NAND FeFET}
\end{flushleft}

\justify
The process integration is performed on a p-type silicon substrate. After phosphorus ion implantation and activation, the isolation oxide in the gate region is removed. RCA clean is then performed to clean the surface of the gate region. The gate dielectric layers, Hf\textsubscript{0.5}Zr\textsubscript{0.5}O\textsubscript{2} and Al\textsubscript{2}O\textsubscript{3}, are deposited through atomic layer deposition (ALD) at the temperatures of 250 \textdegree C and 200 \textdegree C, respectively. Source/drain via is opened by dry etch (reactive-ion etch, RIE) and wet etch (buffered oxide etch, BOE). A 70-nm-thick tungsten (W) layer is sputtered on the sample to serve as source, drain, and gate contact metal. Finally, the sample is annealed through rapid thermal processing (RTP) in forming gas (N\textsubscript{2}+H\textsubscript{2}) at 350 \textdegree C for 1 min and N\textsubscript{2} at 500 \textdegree C for 20 s (Fig.\ref{fig_s1}(a)). Fig.\ref{fig_s1}(b) and (c) show the STEM and elemental map of the gate stack from tungsten to silicon channel region.

\begin{figurehere}
	\centering
	\includegraphics[width=1\linewidth]{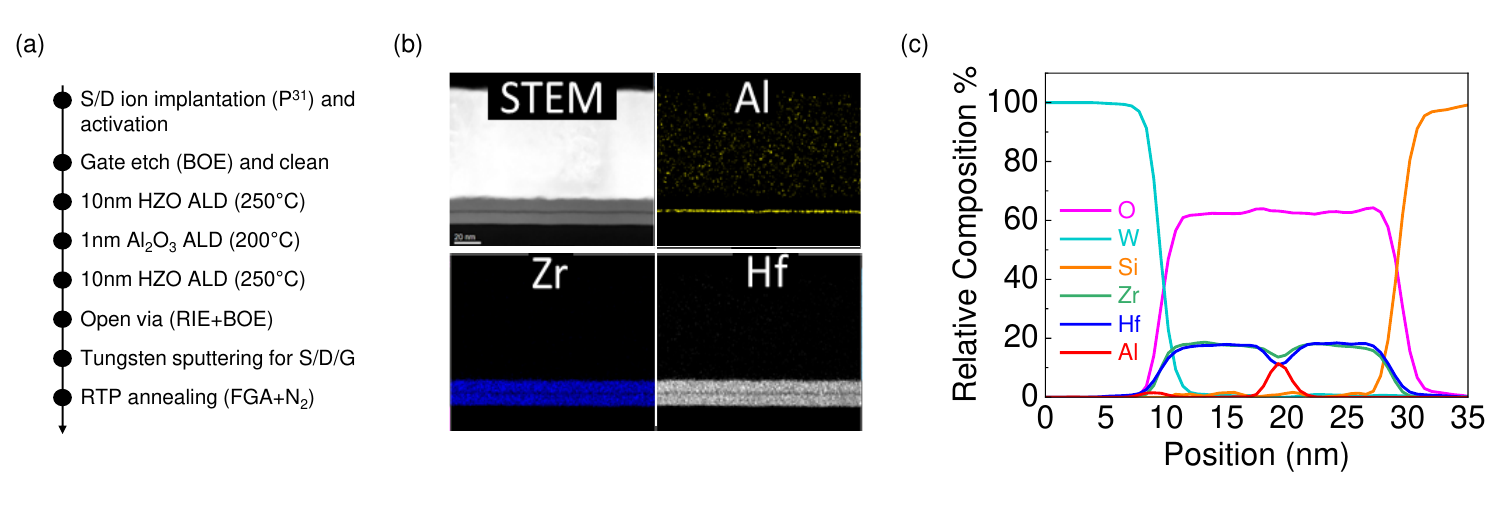}
	\caption{\textit{\textbf{Process integration and gate stack metrology.} (a) Process integration flow. (b) STEM shows different layers in the gate stack. (c) Elemental map along the stack position confirms the Al\textsubscript{2}O\textsubscript{3} layer is inserted into the 20 nm HZO.}}
	\label{fig_s1}
\end{figurehere}

\newpage
\begin{flushleft} 
\textbf{\large FeFET Variation}
\end{flushleft}

The FeFET variation is investigated by measuring ten FeFETs with the same program pulses. The devices are initially reset to \textit{V}\textsubscript{TH0H} state. Then the devices are programmed with different pulses to achieve MLC, as shown in Fig.\ref{fig_s2}(a). No overlap is observed between different \textit{V}\textsubscript{TH} states. The cumulative probability of \textit{V}\textsubscript{TH} distribution is shown in Fig.\ref{fig_s2}(b). These results indicate that MLC is feasible for the proposed application.

\begin{figurehere}
	\centering
	\includegraphics [width=1\linewidth]{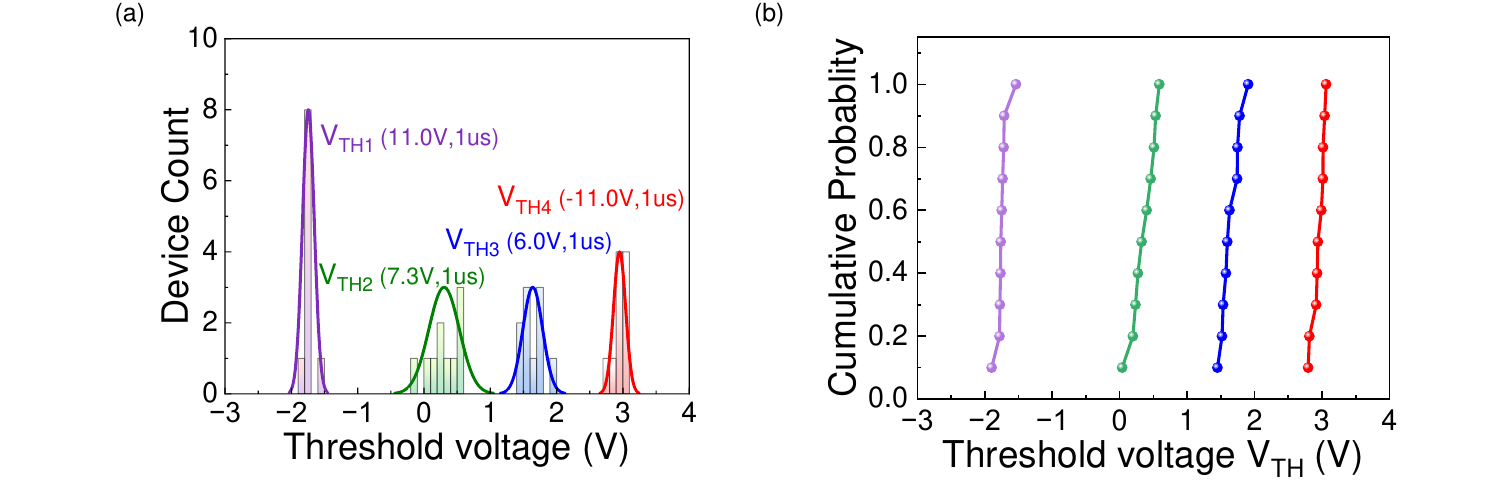}
	\caption{\textbf{FeFET variation. } \textit{(a) MLC state \textit{V}\textsubscript{TH} distribution. (b) Cumulative distribution of  \textit{V}\textsubscript{TH} states.}}
	\label{fig_s2}
\end{figurehere}

\newpage
\begin{flushleft} 
\textbf{\large Spatial-Temporal Sequence Detection}
\end{flushleft}

Additional experimental results are shown in Fig.\ref{fig_s3}. The read cycle is defined as the last cycle to achieve temporal detection. When all WLs are passing, the input can be passed to the output. A high current is detected in the read cycle (Fig.\ref{fig_s3}(a)). Here the fifth FeFET is used as a pass transistor and is always written to low-\textit{V}\textsubscript{TH} state. In Fig.\ref{fig_s3}(b), the input pulses match the stored patterns. A high current is thus detected in the read cycle. In Fig.\ref{fig_s3}(c), a mismatch occurs, resulting in a low mismatch current.  

\begin{figurehere}
	\centering
	\includegraphics[width=1\linewidth]{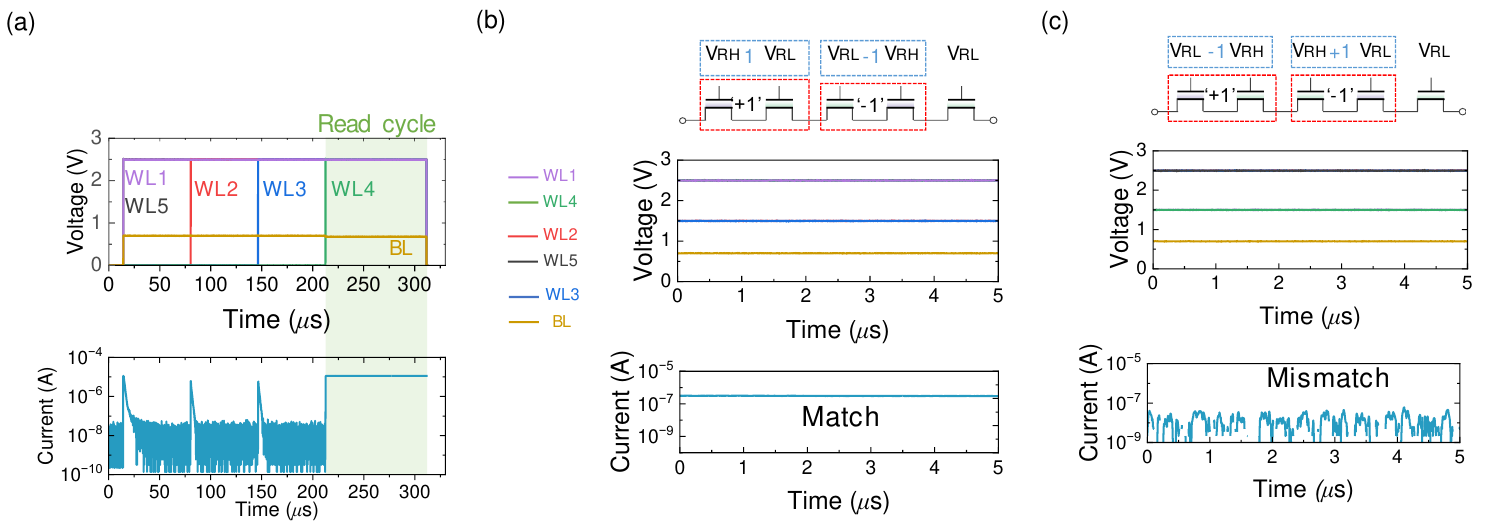}
	\caption{\textit{\textbf{Additional experimental results for the spatial-temporal sequence detection.} (a) Capability of a string in passing the input BL voltage to the output. (b) Readout current of a match condition. (c) Readout current of a mismatch condition.}}
	\label{fig_s3}
\end{figurehere}

\end{document}